\documentclass[10pt,a4paper]{article}
\usepackage[utf8]{inputenc}
\usepackage{lmodern}
\usepackage[T1]{fontenc}
\usepackage{relsize}

\usepackage[margin=3cm,top=2cm,bottom=2.5cm]{geometry}

\usepackage{silence}
\usepackage{microtype}
\usepackage{parskip}
\usepackage[all]{nowidow}

\usepackage{tabularx}
\usepackage{booktabs}
\usepackage[shortlabels]{enumitem}
\usepackage{float}
\usepackage{xspace}

\usepackage{xparse}
\usepackage{listofitems}

\usepackage{amsmath}
\usepackage{mathtools}
\usepackage{amssymb}
\usepackage{tensor}
\usepackage{centernot}
\usepackage{cancel}

\usepackage[useregional,showisoZ=false]{datetime2}
\usepackage{bussproofs}

\usepackage{tikz}
\usetikzlibrary{shapes, arrows, arrows.meta, positioning, decorations.pathmorphing, backgrounds, fit, calc}

\usepackage{xcolor}
\usepackage{color}

\usepackage{ifplatform}
\usepackage{ifthen}

\usepackage[
    backend=biber,
    style=alphabetic,
    natbib=true,
    url=true,
    maxcitenames=50,
    maxbibnames=50,
    abbreviate=false,
    eprint=false,
    doi=true
]{biblatex}
\DeclareFieldFormat{url}{\url{#1}}
\DeclareFieldFormat{doi}{\url{https://doi.org/#1}}
\usepackage{amsthm}
\usepackage{thmtools}
\newtheorem{theorem}{Theorem}

\newtheorem{definition}[theorem]{Definition}
\newtheorem{lemma}[theorem]{Lemma}
\newtheorem{fact}[theorem]{Fact}

\newtheorem{remark}[theorem]{Remark}

\usepackage{hyperref}
\hypersetup{
  colorlinks=true,
  linkcolor=black,
  citecolor=black,
  urlcolor=gray,
  linktoc=all,
  breaklinks=true,
}

\newcommand{\isdef}{\mathrel{:=}}

\newcommand{\eq}{\leftrightarrow}

\newcommand{\pdlforms}{\mathcal{F}}

\newcommand{\kmodel}{\mathcal{M}}
\newcommand{\M}{\mathcal{M}}
\renewcommand{\phi}{\varphi}

\newcommand{\seqc}{\mathop{;}} 

\makeatletter
\newcommand{\BNF}[2]{\begin{equation*}#1 \, \, \Coloneqq \, \, #2\BNFchecknextarg}
\newcommand{\BNFchecknextarg}{\@ifnextchar\bgroup{\BNFgobblenextarg}{\end{equation*}}}
\newcommand{\BNFgobblenextarg}[1]{\, \, | \, \, #1\@ifnextchar\bgroup{\BNFgobblenextarg}{\end{equation*}}}
\makeatother

\newcommand{\rel}[1]{\xrightarrow{#1}}
\newcommand{\reach}[1]{R_{#1}}

\newcommand{\progs}{\mathcal{P}}

\newcommand{\Nats}{\mathbb{N}}

\renewcommand{\models}{\Vdash}

\newcommand{\powerset}{\mathcal{P}}

\newcommand{\boxvec}{\overrightarrow{\Box\strut_{\phantom{x}}\kern-1ex}}

\newcommand{\voc}{\mathsf{voc}}

\newread\fid
\newcommand{\readfile}[1]
{\bgroup
  \endlinechar=-1
  \openin\fid=#1
  \read\fid to\filetext
  \loop\ifx\empty\filetext\relax
    \read\fid to\filetext
  \repeat
  \closein\fid
  \global\let\filetext=\filetext
\egroup}

\usepackage{ifplatform}
\newcommand{\whenlean}[2]{%
  \ifshellescape%
    \IfFileExists{./tools/finder}{#1}{#2}%
  \else%
    #2\fi%
}

\newcommand{\lean}[1]{%
  \href{https://malv.in/2026/test-free-pdl-no-interpolation/search/?q=#1}{\texttt{\detokenize{#1}}}%
}

\newcommand{\depth}{\mathsf{depth}}
\newcommand{\MG}{\ensuremath{\mathcal{M}_\Gamma}}

\newcommand{\pause}{}

\author{Malvin Gattinger \\
  ILLC, University of Amsterdam \\
  \href{mailto:malvin@w4eg.eu}{\texttt{malvin@w4eg.eu}}
}
\date{12 November 2026}
\title{Propositional Dynamic Logic without tests \texorpdfstring{\\}{} does not have Craig Interpolation}

\begin{document}

\maketitle

\begin{abstract}
  It was shown recently by Borzechowski et al.\ that Propositional Dynamic Logic (PDL) has Craig Interpolation.
  Whether test-free PDL, i.e.\ PDL without the `{?}' operator, has interpolation remained open.
  Here we answer this question in the negative with a counterexample:
  a test-free PDL validity for which no test-free interpolant exists.
  The validity only uses one atomic program and the Kleene star, so it implies that the basic modal logic K extended with the reflexive transitive closure does not have interpolation, as previously shown by Studer (2009).

  The example was originally obtained using LLMs and Lean.
  We describe our methodology and provide links to all prompts, LLM output summaries as well as the Lean proof.
\end{abstract}

\section{Introduction}

Propositional Dynamic Logic (PDL), originally from~\cite{FL:PDL1979}, is a modal logic that reasons about the execution of programs.
PDL can express many properties of labelled transition systems, for example ``If we keep executing the program $a$ while $p$ holds, then whenever the loop terminates, $q$ holds'' can be stated in PDL as $[{(p?;a)}^*;\lnot p?]q$.

Recently Borzechowski et al.~\cite{PDLhasCI2026}, using ideas from~\cite{MB1988}, have shown that PDL has Craig Interpolation, meaning that for any implication $\phi \to \psi$ valid in PDL we can find a so-called interpolant $\theta$ that only uses the proposition letters and programs occurring both in $\phi$ and in $\psi$ and such that $\phi \to \theta$ and $\theta \to \psi$ are also valid.
For example, $[a \cup b^*](p \land q) \to [b;b^*;r?]q$ is valid and $[b^*]q$ is an interpolant for it.

The \emph{test-free} fragment of PDL consists of the PDL programs and formulas without the ``{?}'' operator.
In this note we show that test-free PDL does not have interpolation.
That is, we provide a test-free valid implication for which no test-free interpolant exists.

This result was first obtained by prompting LLMs.
This document was written by its human author and is structured as follows.
In \autoref{sec:related} we briefly discuss related work and in \autoref{sec:basics} we recall the basic syntax and semantics of PDL.
Readers already very familiar with PDL may want to jump to \autoref{sec:overview} where we define the validity used as the interpolation counterexample and give an overview of our argument.
In \autoref{sec:semilinear} we make a small detour and discuss how the single-program fragment of test-free PDL relates to semilinear subsets of natural numbers via Parikh's theorem.
Getting back to our main story we then provide the countermodel and the remaining proofs in \autoref{sec:modelAndProofs}.
Lastly, we describe the use of LLMs and Lean in \autoref{sec:methodology}, where we also provide links to all prompts and summaries of the LLM output.

\section{Related Work}\label{sec:related}

It is well-known that the test-free fragment is less expressive than full PDL~\cite{BerPat1981:testFreeWeaker}.
The proof for this uses the formula
  $\langle {(p?; a)}^*; \lnot p? ; a; p? \rangle \top$
which can be rewritten as
  $\langle (\mathsf{while}\ p \ \mathsf{do}\ a); a; p? \rangle \top$
and which also served as inspiration for the interpolation counterexample we present here.

It was shown in~\cite[Theorem~10.6.2]{Kracht1999tools} that interpolation for test-free PDL implies interpolation for full {PDL}.
Note the direction here --- it means that the result from~\cite{PDLhasCI2026} that full PDL has interpolation did \emph{not} yet settle whether test-free PDL has interpolation.

Interpolation has been (dis)proven for many logics, including some related to PDL.
In particular,~\cite{MaksimovaNextNoBeth1991} shows that temporal logics with a ``next'' operator do not have interpolation.
On the other hand, some fragments of LTL do have interpolation~\cite{GheerCateLTLcraig2009}.

The ideas from~\cite{MaksimovaNextNoBeth1991} were later used by~\cite{StuderCKnoBeth2009} to show that multi-modal K with any number of agents, extended with common knowledge for the group of all agents, does not have the Beth property and hence no interpolation, and that the same holds over transitive frames (K4) when there are at least two agents.

\section{Basics}\label{sec:basics}

We recall the main relevant definitions of PDL, using the same notation as~\cite{PDLhasCI2026}.

\begin{definition}\label{d:pdl-syntax}
  Let $\pdlforms_0$ and $\progs_0$ be countably infinite sets of proposition letters and atomic programs, respectively.
  The PDL formulas and PDL programs are defined as follows:
  \BNF{\pdlforms \ni \phi, \psi, \tau}{\bot \mid p \mid \lnot \phi \mid \phi \land \psi \mid [\alpha] \phi}
  \BNF{\progs \ni \alpha, \beta}{a \mid \tau? \mid \alpha \cup \beta \mid \alpha ; \beta \mid \alpha^\ast}
  where $p \in \pdlforms_0$ and $a \in \progs_0$.
  Other Boolean connectives $\top$, $\lor$, $\to$ and $\eq$ are defined as usual in terms of $\neg$ and $\land$.
  The diamond modality is defined as $\langle\alpha\rangle\phi := \lnot[\alpha]\lnot\phi$.

  A formula or a program is called \emph{test-free} iff it does not contain the `{?}' operator.
\end{definition}

\begin{definition}\label{d:vocab}
Let $\xi$ be a formula or a program.
The vocabulary~$\voc(\xi)$ of $\xi$ is defined as the set of proposition letters and atomic programs occurring in $\xi$.
Moreover, for any set of formulas $\Gamma$, we define $\voc(\Gamma) \isdef  \bigcup \{ \voc(\phi) \mid \phi \in \Gamma \}$.
\end{definition}
For example, we have $\voc([{(p? \cup a)}^*;b](q \lor \lnot r)) = \{a,b,p,q,r\}$.

\begin{definition}\label{d:KripkeModel}
  A \emph{Kripke model} $\kmodel = (W, \{\reach{a} \mid a \in \progs_0\}, V)$ consists of a set $W$ of states, an accessibility relation $\reach{a} \subseteq W^2$ for every atomic program $a \in \progs_0$, and a valuation $V \colon \pdlforms_0 \to \powerset(W)$ that assigns to each proposition letter a subset of $W$.
  A~pointed model is a pair $(\kmodel, v)$ where $\kmodel$ is a Kripke model and $v$ is a state in $\kmodel$.
\end{definition}

We denote the natural numbers $\{0,1,2,\ldots\}$ by $\Nats$.

\begin{definition}\label{d:evaluate}
  Let $\kmodel= (W, \{\reach{a} \mid a \in \progs_0\}, V)$ be a Kripke model.
  We interpret formulas and programs on $\kmodel$ by a mutual induction as follows:
  \[
    \begin{array}{lll}
      \kmodel, v \models \bot            & \text{ iff } & \text{never}                                                               \\
      \kmodel, v \models p               & \text{ iff } & v \in V(p)                                                                 \\
      \kmodel, v \models \lnot \phi      & \text{ iff } & \kmodel, v \not\models \phi                                                \\
      \kmodel, v \models \phi \land \psi & \text{ iff } & \kmodel, v \models \phi \text{ and } \kmodel, v \models \psi               \\
      \kmodel, v \models [\alpha] \phi   & \text{ iff } & \kmodel, w \models \phi \text{ for all $w$ with } (v,w) \in \reach{\alpha} \\[1em]
    \end{array}
  \]
  \vspace{-1em}
  \[
    \begin{array}{lll}
      \reach{\phi?}             & \isdef & \{ (v,v) \mid v \models \phi \}  \hfill (\text{test})                                                                             \\
      \reach{\alpha \cup \beta} & \isdef & \reach{\alpha} \cup \reach{\beta}  \hfill (\text{choice})                                                                         \\
      \reach{\alpha ; \beta}    & \isdef & R_\alpha \seqc R_\beta = \{(w,v) \mid \exists u: (w,u) \in R_\alpha \land (u,v) \in R_{\beta}\}  \;\; \hfill (\text{composition}) \\
      \reach{\alpha^\ast}       & \isdef & \reach{\alpha}^\ast = \bigcup_{n \in \Nats} R_\alpha^n\hfill (\text{reflexive transitive closure})
    \end{array}
  \]
  For $(v, w) \in \reach{\alpha}$ we also write $v \rel{\alpha} w$ and we also write $w \rel{a^n} v$ for $(w,v) \in R_a^n$.
  Given a set $X$ of formulas, we write $\kmodel, v \models X$ if $\kmodel, v \models \phi$ for all $\phi \in X$.
  We say that a formula is \emph{valid} iff we have $\kmodel, v \models \phi$ for all pointed models $(\kmodel,v)$.
  We will often just write $v \models \phi$ instead of $\M, v \models \phi$ when $\M$ is clear from the context.
\end{definition}

\begin{definition}[\lean{Interpolant}]\label{d:Interpolant}
Let $\phi$ and $\psi$ be PDL formulas such that $\phi \to \psi$ is valid.
An \emph{interpolant} for $\phi \to \psi$ is a formula $\theta$ such that
$\voc(\theta) \subseteq \voc(\phi) \cap \voc(\psi)$,
$\models \phi \to \theta$, and
$\models \theta \to \psi$ hold.

Let $\Gamma_1$ and $\Gamma_2$ be finite sets of formulas such that $\Gamma_1 \cup \Gamma_2$ is not satisfiable.
If $\theta$ is an interpolant for $\bigwedge\Gamma_1 \to \lnot\bigwedge\Gamma_2$ then we also call $\theta$ an interpolant for $\Gamma_1;\Gamma_2$.
\end{definition}

The proof we will give in this note does not rely on any tableaux or other proof system, so we do not need any other definitions from~\cite{PDLhasCI2026}.

We will use the following definition of modal depth.
It is shallow in the sense that it does not care about the structure of programs.
For example, $\langle a \cup b \rangle p$ and $\langle a ; (b^* \cup c) \rangle p$ both have depth 1.

\begin{definition}[\lean{mdepth}]\label{def:depth}
  For any test-free formula $\phi$ we define its \emph{modal depth} as follows.
  \[
    \begin{array}{lll}
      \depth(\bot)            & := & 0 \\
      \depth(p)               & := & 0 \\
      \depth(\lnot \phi)      & := & \depth(\phi) \\
      \depth(\phi \land \psi) & := & \max(\depth(\phi),\depth(\psi)) \\
      \depth([\alpha] \phi)   & := & \depth(\phi) + 1 \\
    \end{array}
  \]
\end{definition}

\begin{definition}[\lean{agree}]\label{def:agree}
  We say that any two states $w_1$ and $w_2$ \emph{agree test-free on $A$ up to $j$}
  and write $w_1 \equiv^A_j w_2$ iff
  for all test-free formulas $\phi$ with $\voc(\phi) \subseteq A$ and $\depth(\phi) \leq j$
  we have $w_1 \models \phi$ iff $w_2 \models \phi$.
\end{definition}

\section{Counterexample and Overview}\label{sec:overview}

For the whole article, we fix two sets of formulas:
\[ \Gamma_1 := \{ s,\ [a^*]\big((p \land s) \to (\langle a\rangle\top \land [a] s)\big) \,\} \]
\[ \Gamma_2 := \{ \ [a^*](q \to p), \ [a^*](\langle a \rangle q \to q), \ \langle a^* \rangle(q \land [a]\bot) \} \]

Note that the common vocabulary is $\voc(\Gamma_1) \cap \voc(\Gamma_2) = \{a,p\}$ (see \lean{voc_gammaOne_gammaTwo}).

We will proceed as follows:
\begin{enumerate}
\item[(1)] We show that $\Gamma_1 \cup \Gamma_2$ is not satisfiable and thus $\bigwedge\Gamma_1 \to \lnot \bigwedge\Gamma_2$ is valid.~(\autoref{l:unsat})
\item[(2)] We define a Kripke model $\MG$ (in \autoref{d:MG}) such that
  $\MG, m \models \Gamma_1$ and $\MG, v_j \models \Gamma_2$ for all $j \in \Nats$.~(\autoref{l:modelSatGammaOneTwo})
\item[(3)] Suppose $\theta$ is an interpolant for $\Gamma_1; \Gamma_2$.
  Then $\Gamma_1 \cup \{\lnot \theta\}$ and $\Gamma_2 \cup \{\theta\}$ are both not satisfiable.
  Moreover, suppose that $\theta$ is test-free.
\item[(4)] Let $j$ be the modal depth of $\theta$ as in \autoref{def:depth} above.
\item[(5)] We show that in $\MG$ the states $m$ and $v_j$ agree on all test-free $\{a,p\}$-formulas of modal depth at most $j$.~(\autoref{l:depthEquiv})
\item[(6)] By (2) and (3) state $m$ must satisfy $\theta$ and state $v_j$ must satisfy $\lnot \theta$, contradicting (5).
\end{enumerate}

We will then be able to conclude that there is no interpolant $\theta$ for $\Gamma_1;\Gamma_2$.

Let us start by showing that the two sets of formulas are indeed not jointly satisfiable.

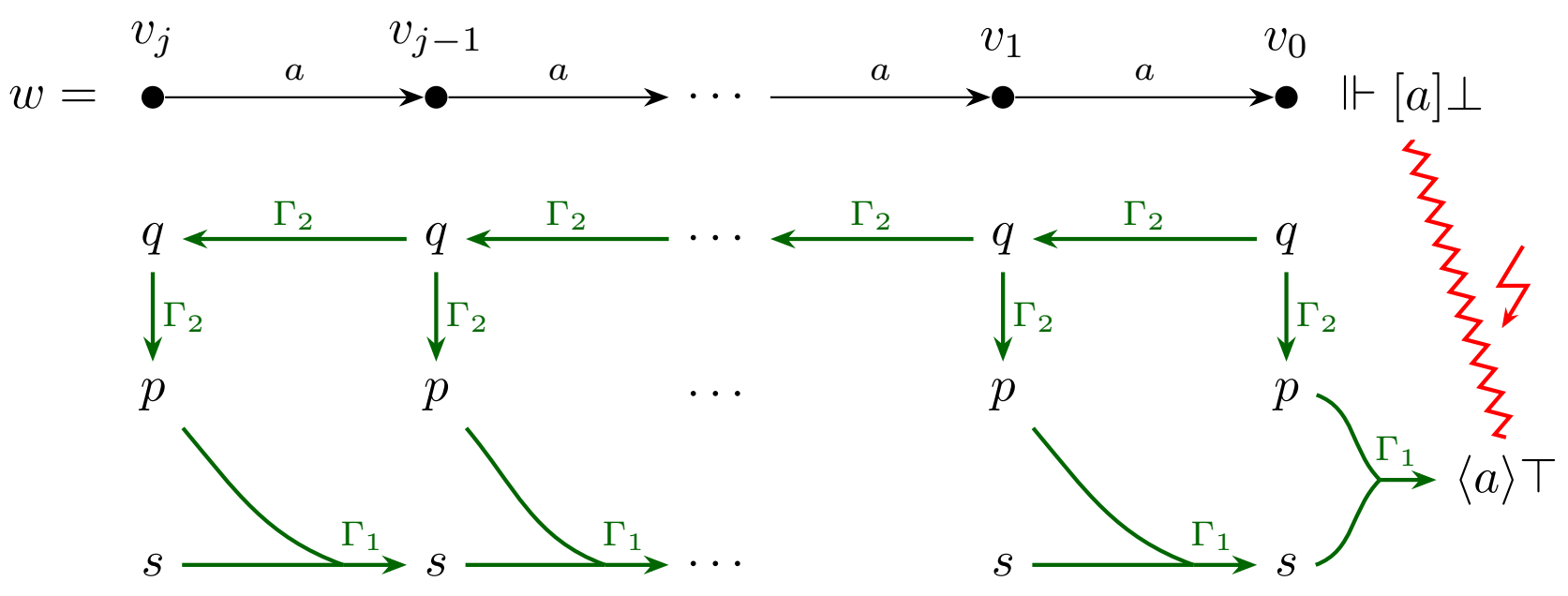
\begin{figure}[ht]
  \centering
  \begin{tikzpicture}[
  >={Stealth[length=5pt]},
  st/.style={circle,fill,inner sep=1.6pt},
  by/.style={->,green!40!black,thick},
  lab/.style={font=\scriptsize,green!40!black,inner sep=1.5pt},
  zz/.style={red,thick,decorate,decoration={zigzag,segment length=5pt,amplitude=2pt}}
  ]
  \foreach \i/\n in {0/j, 1/j-1, 3/1, 4/0} {
    \node[st,label={[label distance=2pt]above:{$v_{\n}$}}] (v\i) at (2*\i,0) {};
  }
  \node (vd) at (4,0) {$\cdots$};
  \node[anchor=east] at ([xshift=-5pt]v0.west) {$w =$};
  \draw[->] (v0) -- node[above,font=\scriptsize] {$a$} (v1);
  \draw[->] (v1) -- node[above,font=\scriptsize] {$a$} (vd);
  \draw[->] (vd) -- node[above,font=\scriptsize] {$a$} (v3);
  \draw[->] (v3) -- node[above,font=\scriptsize] {$a$} (v4);
  \node[anchor=west] (abot) at ([xshift=5pt]v4.east) {$\models [a]\bot$};
  \node (q4) at (2*4,-1.0) {$q$};
  \pause

  \node (q3) at (2*3,-1.0) {$q$};
  \draw[by] (q4) -- node[lab,above] {$\Gamma_2$} (q3);
  \pause
  \node (qd) at (4,-1.0) {$\cdots$};
  \draw[by] (q3) -- node[lab,above] {$\Gamma_2$} (qd);
  \pause
  \node (q1) at (2*1,-1.0) {$q$};
  \draw[by] (qd) -- node[lab,above] {$\Gamma_2$} (q1);
  \pause
  \node (q0) at (2*0,-1.0) {$q$};
  \draw[by] (q1) -- node[lab,above] {$\Gamma_2$} (q0);
  \pause

  \foreach \i in {0,1,3,4} {
    \node (p\i) at (2*\i,-2.1) {$p$};
    \draw[by] (q\i) -- node[lab,right] {$\Gamma_2$} (p\i);
  }
  \node (pd) at (4,-2.1) {$\cdots$};
  \pause


  \newcommand{\sstep}[3]{
    \coordinate (m) at ($(#3.west)+(-0.45,0)$);
    \draw[green!40!black,thick] (#1.south east) to[out=-50,in=160] (m);
    \draw[green!40!black,thick] (#2.east) -- (m);
    \draw[by] (m) -- node[lab,above,pos=0.3,yshift=1pt] {$\Gamma_1$} (#3.west);
  }

  \node (s0) at (2*0,-3.3) {$s$};
  \node (s1) at (2*1,-3.3) {$s$};
  \sstep{p0}{s0}{s1}
  \pause
  \node (sd) at (4,-3.3) {$\cdots$};
  \sstep{p1}{s1}{sd}
  \pause
  \node (s3) at (2*3,-3.3) {$s$};
  \node (s4) at (2*4,-3.3) {$s$};
  \sstep{p3}{s3}{s4}
  \pause

  \node (aT) at (9.55,-2.7) {$\langle a\rangle\top$};
  \coordinate (mT) at ($(aT.west)+(-0.4,0)$);
  \draw[green!40!black,thick] (p4.east) to[out=-20,in=135] (mT);
  \draw[green!40!black,thick] (s4.east) to[out=20,in=-135] (mT);
  \draw[by] (mT) -- node[lab,above,pos=0.3,yshift=1pt] {$\Gamma_1$} (aT.west);
  \pause
  \draw[zz] (aT.north) -- (abot.south);
  \coordinate (zm) at ($(aT.north)!0.5!(abot.south)+(0.45,0.3)$);
  \draw[red,thick,-{Stealth[length=4pt]}] (zm) -- ++(-0.17,-0.28) -- ++(0.2,0) -- ++(-0.18,-0.3);
\end{tikzpicture}
  \caption{Illustration of the proof of \autoref{l:unsat}.}\label{fig:unsat}
\end{figure}

\begin{lemma}[\lean{gammaOne_gammaTwo_unsat}, \lean{gammaOne_gammaTwo_taut}]\label{l:unsat}
  The set $\Gamma_1 \cup \Gamma_2$ is not satisfiable
  and thus $\bigwedge\Gamma_1 \to \lnot \bigwedge\Gamma_2$ is valid.
\end{lemma}
\begin{proof}
  One can use any sound and complete proof system for PDL to show this.
  The system from~\cite{PDLhasCI2026} with its Haskell implementation also yields an interpolant,
  a rather long formula equivalent to $[{(p?;a)}^*](p \to \langle a \rangle\top)$.
  As expected, this formula uses a test.
  We still illustrate the contradiction between $\Gamma_1$ and $\Gamma_2$ with a semantic argument below.

  In order to reach a contradiction, suppose there exists a Kripke model $\M$ with a state $w$ such that $\M,w \models \Gamma_1 \cup \Gamma_2$.
  We illustrate the following reasoning in \autoref{fig:unsat}.

  By $w \models \langle a^* \rangle (q \land [a]\bot)$ from $\Gamma_2$
  there exists a $j \in \Nats$ and a chain
  $w =: v_j \rel{a} v_{j-1} \rel{a} \ldots \rel{a} v_0 \models q \land [a]\bot$.

  By using $w \models [a^*](\langle a \rangle q \to q)$ backwards along this chain we get $v_i \models q$ for all $i \leq j$.
  In addition from $w \models [a^*](q \to p)$ (from $\Gamma_2$) we get $v_i \models p$ for all $i \leq j$.

  Now from $w = v_j \models p \land s$ (using $s \in \Gamma_1$) we get from $\Gamma_1$ that $v_j \models [a]s$ and therefore $v_{j-1} \models s$.
  Iterating this last argument forward along the chain we get $v_i \models s$ for all $i \leq j$.

  But now we have $v_0 \models p \land s$ and then from $\Gamma_1$ get $v_0 \models \langle a \rangle \top$ which contradicts $v_0 \models [a]\bot$.
\end{proof}

\section{Single-program Test-free PDL and Semilinear Sets}\label{sec:semilinear}

In $\Gamma_1 \cup \Gamma_2$ we only use a single atomic program $a$.
It turns out that in such a single-program fragment of PDL we can rephrase the semantics of $[\alpha]\phi$ for any test-free $\alpha$ as follows.
At $w$ the formula $[\alpha]\phi$ is true iff for any $n \in \Nats$ such that $\alpha$ can be executed by making $n$ many $a$-steps we have that at any state reachable in $n$ steps $\phi$ holds. (This will be \autoref{l:boxIffLengths}.)

Let us make this precise by defining the set of execution lengths.

\begin{definition}[\lean{L}]\label{d:excLen}
  For any test-free program $\alpha$ such that $\voc(\alpha)=\{a\}$,
  we define its set of \emph{possible execution lengths} $L(\alpha)$:
  \[
    \begin{array}{lcl}
      L(a) & := & \{ 1 \} \\
      L(\alpha ; \beta) & := & \{ n_\alpha + n_\beta \mid n_\alpha \in L(\alpha), n_\beta \in L(\beta) \} \\
      L(\alpha \cup \beta) & := & L(\alpha) \cup L(\beta) \\
      L(\alpha^*) & := & \bigcup_{l \geq 0} \{ \sum_{0 \leq i < l} n_i \mid n_0, \ldots, n_{l-1} \in L(\alpha) \} \\ 
    \end{array}
  \]
\end{definition}

Given this definition we can also explain the title of this section.
Our definition of $L$ is what has been called the Parikh image of the regular language given by $\alpha$ over a single-character alphabet.
The \emph{semilinear subsets}~\cite{GinsburgSpanier1966} of $\Nats$ are formed by taking finite unions of sets of the form
$\{ c + \lambda_1 p_1 + \ldots + \lambda_m p_m \mid \lambda_i \in \Nats \}$
where $c, p_1, \ldots, p_m \in \Nats$.
The following fact is then a special case of Parikh's theorem.
We refer to~\cite[Theorem~2, p.~573]{Parikh1966} for the original proof and~\cite[Theorem~H.1, p.~202]{Kozen1997} for a textbook explanation.

\begin{fact}[\lean{L_isSemilinearSet}]\label{l:testFreeSemiLinear}
  For any test-free $\alpha$ with $\voc(\alpha) = \{a\}$ the set $L(\alpha)$ is semilinear.
\end{fact}

We note that in the Lean formalization \autoref{l:testFreeSemiLinear} is proved not via Parikh's theorem (which is about context-free languages) but directly by induction on $\alpha$, using the facts from Mathlib that semilinear sets are closed under the three operations used in \autoref{d:excLen}.

\begin{remark}
The converse of \autoref{l:testFreeSemiLinear} does not hold, because every test-free program $\alpha$ with $\voc(\alpha) = \{a\}$ admits some positive number of steps.
So neither $\varnothing$ nor $\{0\}$ is of the form $L(\alpha)$, even though both are semilinear.
Moreover, if $0 \in L(\alpha)$ then $\alpha$ must contain a star, and $L(\alpha)$ then contains all multiples of some period $p > 0$; hence for example $\{0,2\}$ is not of this form either, although it is semilinear.
One can prove that these are the only kinds of exceptions: the sets $L(\alpha)$ given by any test-free program $\alpha$ with $\voc(\alpha) = \{a\}$ are exactly the nonempty semilinear sets $X \subseteq \Nats$ such that $0 \in X$ implies $p\Nats \subseteq X$ for some $p > 0$ (see \lean{exists_L_eq_iff}).

A second way (not formalized yet) to turn \autoref{l:testFreeSemiLinear} into an ``iff'' characterization is to add the $\mathsf{skip}$ program to the program syntax, with the interpretation that $R_\mathsf{skip}$ is the identity.
In full PDL we can define $\mathsf{skip} := \top?$, but in test-free PDL we would need to add it as a primitive.
We would then have $\voc(\mathsf{skip}) = \varnothing$ and set $L(\mathsf{skip}) := \{0\}$.
Then the test-free PDL programs $\alpha$ with $\voc(\alpha) \subseteq \{a\}$ describe exactly the nonempty semilinear sets of natural numbers.
\end{remark}

Let us turn back to PDL semantics now.
With the set $L$ we can characterize the $R_\alpha$-relation and the semantic meaning of $[\alpha]\phi$ as done in the following two lemmas.

\begin{lemma}[\lean{relate_iff}]\label{l:relateIffSomeLength}
  Take any test-free program $\alpha$ such that $\voc(\alpha)=\{a\}$, and any Kripke model $\M$.
  For any two states $w$ and $v$ in $\M$
  we have $w \rel{\alpha} v$ iff there is some $n \in L(\alpha)$ such that $w \rel{a^n} v$.
\end{lemma}
\begin{proof}
  By induction on the structure of the program $\alpha$.
  \begin{itemize}
  \item For $a$ the claim follows directly from the semantics and \autoref{d:excLen}.
  \item For $\alpha;\beta$ we get the following chain of equivalences:
    \[
      \begin{array}{lll}
        \    & w \rel{\alpha;\beta} v \\[0.2em]
        \iff & \text{there is a $u_0$ s.t.\ } w \rel{\alpha} u_0 \rel{\beta} v & \text{(by semantics)} \\
        \iff &\text{there are $u_0$, $n_\alpha \in L(\alpha)$, $n_\beta \in L(\beta)$ s.t.\ }
               w \rel{a^{n_\alpha}} u_0 \rel{a^{n_\beta}} v & \text{(by IH for $\alpha$ and $\beta$)} \\
        \iff &\text{there is an $n \in L(\alpha;\beta)$ s.t.\ }
               w \rel{a^{n}} v & \text{(by \autoref{d:excLen})}
      \end{array}
    \]
  \item For $\alpha \cup \beta$ we get the following chain of equivalences:
    \[
      \begin{array}{lll}
        \    & w \rel{\alpha \cup \beta} v \\[0.2em]
        \iff & w \rel{\alpha} v \text{ or } w \rel{\beta} v & \text{(by semantics)} \\
        \iff & \text{there is an $n \in L(\alpha)$ s.t.\ } w \rel{a^n} v \text{ or }\\
        \    & \text{there is an $n \in L(\beta)$ s.t.\ } w \rel{a^n} v
             & \text{(by IH for $\alpha$ and $\beta$)} \\
        \iff & \text{there is an $n \in L(\alpha \cup \beta)$ s.t.\ }
               w \rel{a^n} v & \text{(by \autoref{d:excLen})}
      \end{array}
    \]
  \item
    For $\alpha^*$ we first show left-to-right, so assume $w \rel{\alpha^*} v$.
    We want to find some $n \in L(\alpha^*)$ such that $w \rel{a^n} v$.
    By definition of $\rel{\alpha^*}$ there exists a chain of states
    $w = u_0 \rel{\alpha} u_1 \rel{\alpha} \ldots \rel{\alpha} u_l = v$ for some $l \geq 0$.
    By IH for $\alpha$ this means there exist $n_0, \ldots, n_{l-1} \in L(\alpha)$ such that
    $w = u_0 \rel{a^{n_0}} u_1 \rel{a^{n_1}} \ldots \rel{a^{n_{l-1}}} u_l = v$.
    Now let $n := \sum_{0 \leq i < l} n_i$.
    We have $n \in L(\alpha^*)$ by \autoref{d:excLen}
    and $w \rel{a^n} v$, so we are done.
    \smallskip

    For right-to-left, take any $n \in L(\alpha^*)$ such that $w \rel{a^n} v$.
    We want to show $w \rel{\alpha^*} v$.
    By \autoref{d:excLen} there exist $l \geq 0$ and $n_0, \ldots, n_{l-1} \in L(\alpha)$ such that $n = \sum_{0 \leq i < l} n_i$.
    Then there exists a chain of states such that
    $w = u_0 \rel{a^{n_0}} u_1 \rel{a^{n_1}} \ldots \rel{a^{n_{l-1}}} u_l = v$.
    By IH for $\alpha$ this means we have
    $w = u_0 \rel{\alpha} u_1 \rel{\alpha} \ldots \rel{\alpha} u_l = v$
    and therefore also $w \rel{\alpha^*} v$.
    \qedhere
  \end{itemize}
\end{proof}

\begin{lemma}[\lean{box_iff}]\label{l:boxIffLengths}
  Take any test-free program $\alpha$ such that $\voc(\alpha)=\{a\}$, and any Kripke model $\M$.
  For any state $w$
  we have
  \begin{center}
  $\M, w \models [\alpha]\phi$ \ iff \ %
  for all $n \in L(\alpha)$ and
  for all $v$ such that $w \rel{a^n} v$
  we have $\M, v \models \phi$.
\end{center}
\end{lemma}
\begin{proof}
  We get the following chain of equivalences:
    \[
      \begin{array}{lll}
        \    & w \models [\alpha]\phi \\[0.2em]
        \iff & \text{for all $v$ s.t.\ $w \rel{\alpha} v$
               we have $v \models \phi$}            & \text{(by semantics)} \\
        \iff & \text{for all $v$ s.t.\ }
               \text{there is some $n \in L(\alpha)$ and $w \rel{a^n} v$} \\
        \    & \text{we have } v \models \phi       & \text{(by \autoref{l:relateIffSomeLength})} \\
        \iff & \text{for all $n \in L(\alpha)$, }
               \text{for all $v$ s.t.\ $w \rel{a^{n}} v$
               we have $v \models \phi$}            & \text{(by first-order reasoning)}
      \end{array}
    \]
    \par \vspace{-2.2\baselineskip}
    \qedhere
\end{proof}

\section{Countermodel and Proofs}\label{sec:modelAndProofs}

We now define a specific Kripke model that we will use to show that there is no test-free interpolant for $\Gamma_1;\Gamma_2$.

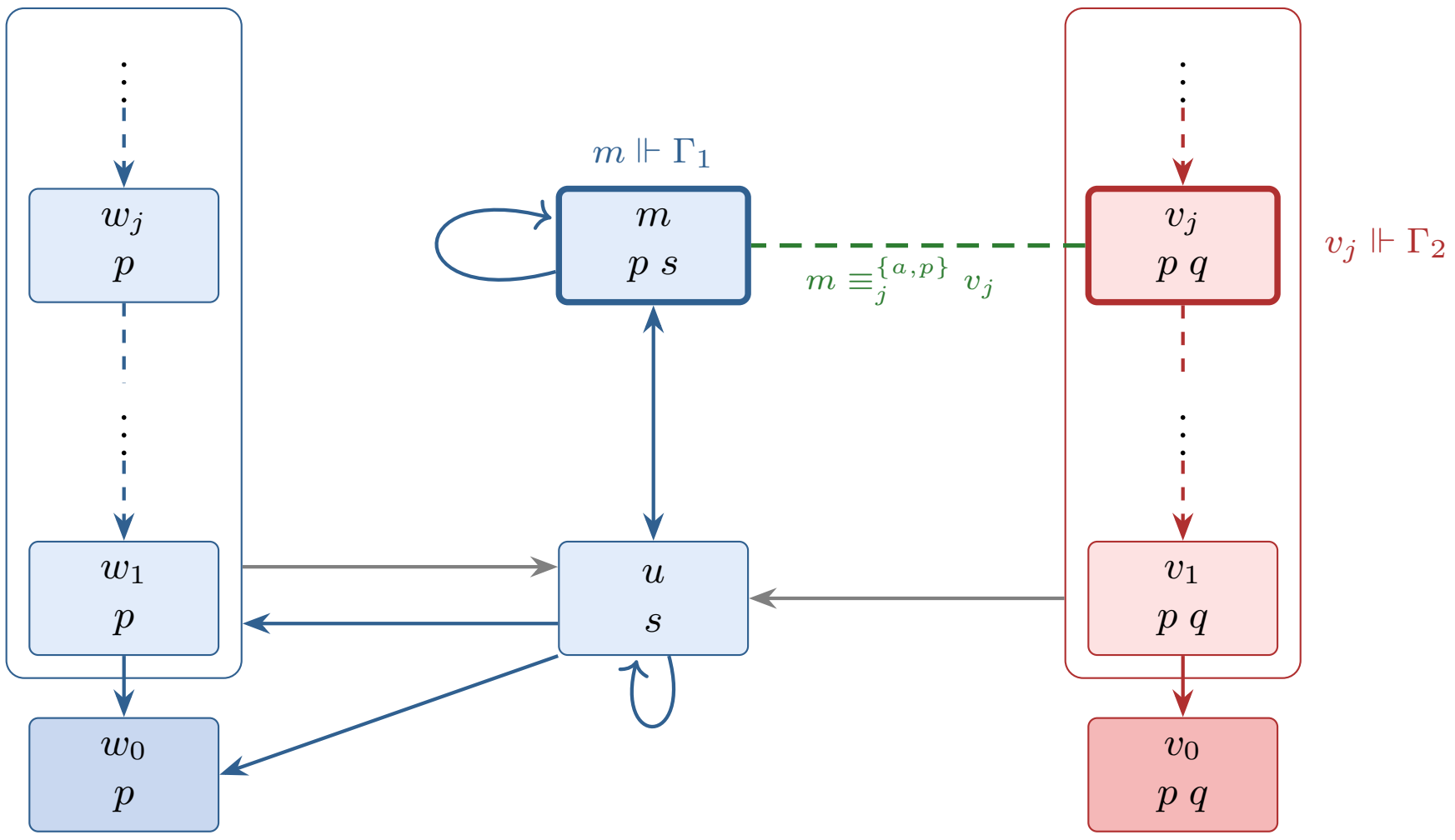
\begin{figure}[ht]
  \centering
\definecolor{mblue}{HTML}{306090}      
\definecolor{mbluefill}{HTML}{E2ECFA}
\definecolor{mbluedark}{HTML}{C9D8F0}
\definecolor{mred}{HTML}{B03030}       
\definecolor{mredfill}{HTML}{FDE2E2}
\definecolor{mreddark}{HTML}{F5B8B8}
\definecolor{mgreen}{HTML}{2E7D32}     

\tikzset{
  state/.style = {draw, rounded corners = 2pt, align = center, font = \small,
                  minimum width = 1.5cm, minimum height = 0.9cm, inner sep = 2pt},
  bstate/.style = {state, draw = mblue, fill = mbluefill},
  bstate dark/.style = {bstate, fill = mbluedark},
  rstate/.style = {state, draw = mred, fill = mredfill},
  rstate dark/.style = {rstate, fill = mreddark},
  marked/.style = {line width = 1.4pt}, 
  dots/.style = {align = center, inner sep = 1pt, minimum height = 0.5cm},
  chain box/.style = {draw, rounded corners = 4pt, inner sep = 5pt},
  arr/.style = {-{Stealth[length = 2.2mm]}, mblue, line width=0.8pt},
  rarr/.style = {-{Stealth[length = 2.2mm]}, mred, line width=0.8pt},
  garr/.style = {-{Stealth[length = 2.2mm]}, gray, line width=0.8pt},
  equiv/.style = {mgreen, dash pattern = on 5pt off 3pt, line width = 1pt},
  annot/.style = {font = \footnotesize, inner sep = 1pt},
}

\begin{tikzpicture}[x = 1cm, y = 1cm]

\node[dots]            (wtop)  at (0, 4.8)    {$\vdots$};
\node[bstate]          (wj)    at (0, 3.4)    {$w_j$\\$p$};
\node[dots]            (wdots) at (0, 2.0)    {$\vdots$};
\node[bstate]          (w1)    at (0, 0.6)    {$w_1$\\$p$};
\node[bstate dark]     (w0)    at (0, -0.8)   {$w_0$\\$p$};
\node[bstate, marked]  (m)     at (4.2, 3.4)  {$m$\\$p\;s$};
\node[bstate]          (hub)   at (4.2, 0.6)  {$u$\\$s$};

\node[dots]            (vtop)  at (8.4, 4.8)  {$\vdots$};
\node[rstate, marked]  (vj)    at (8.4, 3.4)  {$v_j$\\$p\;q$};
\node[dots]            (vdots) at (8.4, 2.0)  {$\vdots$};
\node[rstate]          (v1)    at (8.4, 0.6)  {$v_1$\\$p\;q$};
\node[rstate dark]     (v0)    at (8.4, -0.8) {$v_0$\\$p\;q$};

\begin{scope}[on background layer]
  \node[chain box, draw = mred]  (Cv) [fit = (vtop) (v1)] {};
  \node[chain box, draw = mblue] (Cw) [fit = (wtop) (w1)] {};
\end{scope}

\path (m) edge[arr, loop left] (m);
\draw[arr, {Stealth[length = 2.2mm]}-{Stealth[length = 2.2mm]}] (m) -- (hub);
\path (hub) edge[arr, loop below] (hub);
\draw[arr]  (hub.west |- 0, 0.4) -- (Cw.east |- 0, 0.4);
\draw[garr] (Cw.east |- 0, 0.85) -- (hub.west |- 0, 0.85);
\draw[arr]  (hub.south west) -- (w0.east);
\draw[garr] (Cv.west |- 0, 0.6) -- (hub.east |- 0, 0.6);
\draw[rarr, dashed]    (vtop) -- (vj);
\draw[rarr, dashed, -] (vj) -- (vdots);
\draw[rarr, dashed]    (vdots) -- (v1);
\draw[rarr]            (v1) -- (v0);
\draw[arr, dashed]     (wtop) -- (wj);
\draw[arr, dashed, -]  (wj) -- (wdots);
\draw[arr, dashed]     (wdots) -- (w1);
\draw[arr]             (w1) -- (w0);

\node[annot, mblue, above = 3pt of m] {$m \models \Gamma_1$};
\node[annot, mred, anchor = west] at ([xshift = 4pt]Cv.east |- vj)
  {$v_j \models \Gamma_2$};
\draw[equiv] (vj.west) -- (m.east)
  node[pos = 0.55, below = 1pt, annot, mgreen, font = \scriptsize]
    {$m \equiv^{\{a,p\}}_{j} v_j$};

\end{tikzpicture}
  \caption{The Kripke model $\MG$.
    All arrows are for the same $R_a$ relation.
    An arrow to or from a region means that all states in the region are related.
    The green dashed line illustrates the claim of \autoref{l:depthEquiv}.
    States $w_0$ and $v_0$ with darker shades are the only dead ends.
  }\label{fig:model}
\end{figure}

\begin{definition}[\lean{MG}]\label{d:MG}
  Let $\pdlforms_0 = \{ p, q, s \}$ and $\progs_0 = \{ a \}$.
  We define the model $\MG$ as follows, shown in \autoref{fig:model}.
  \[ W \ := \ \{ m, u \} \cup \{ v_i \mid i \in \Nats \} \cup \{ w_i \mid i \in \Nats \} \]
  \[ R_a \ := \
    \begin{array}{ll}
      \    & \{ (m,m), (m,u), (u,m), (u,u) \} \\
      \cup & \{ (w_i, w_{i-1}) \mid 0 < i \}
             \cup \{ (w_i, u) \mid 0 < i \} \\
      \cup & \{ (u, w_i) \mid 0 \leq i \} \\
      \cup & \{ (v_i, v_{i-1}) \mid 0 < i \}
             \cup \{ (v_i, u) \mid 0 < i \}
    \end{array}
  \]
  \[
    \begin{array}{lcl}
      V(p) & := & \{ m \} \cup \{ w_i, v_i \mid i \in \Nats \} \\
      V(q) & := & \{ v_i \mid i \in \Nats \} \\
      V(s) & := & \{ m, u \} \\
    \end{array}
  \]
\end{definition}

Next we will check that in the model $\MG$ the two sets are satisfied at two different states.
\begin{lemma}[\lean{model_sat}]\label{l:modelSatGammaOneTwo}
  In model $\MG$ from \autoref{d:MG} we have $m \models \Gamma_1$ and $v_j \models \Gamma_2$ for any $j \in \Nats$.
\end{lemma}
\begin{proof}
  We check each of the five formulas:
  \begin{itemize}
  \item $m \models s$ is obvious.
  \item $m \models [a^*]\big((p \land s) \to (\langle a\rangle\top \land [a] s)\big)$ holds because the only state that is $a^*$ reachable from $m$ and satisfies $p \land s$ is the state $m$ itself, and $m$ has an $a$-successor and all its $a$-successors (which are $u$ and itself) satisfy $s$.
  \item $v_j \models [a^*](q \to p)$ is easy to see because $q \to p$ is globally true in the model.
  \item $v_j \models [a^*](\langle a \rangle q \to q)$ holds because any state with a $q$-successor is also a $q$-state.
  \item $v_j \models \langle a^* \rangle(q \land [a]\bot)$ is witnessed by the $a$-chain to $v_0$ (on the right in red in \autoref{fig:model}.)
    \qedhere
  \end{itemize}
\end{proof}

We now prove a helper lemma, showing that the $w_i$- and the $v_i$-chains in $\MG$ agree on all $\{a,p\}$-formulas.
Intuitively, this is because each $v_i$ and $w_i$ have the same valuation when restricted to $p$,
and because the relations among the $v_i$ are the same as among the $w_i$.
The only difference between the chains is that from $u$ we can reach all $w_i$ but no $v_i$.
But this does not affect which states can be reached \emph{from} each chain.
We make this statement and argument precise in the following lemma and its proof using a bisimulation.

\begin{lemma}[\lean{eval_w_v}]\label{l:chainBisim}
  In the model $\MG$ we have
  for all $i \in \Nats$ and all $\{a,p\}$-formulas $\phi$ that $v_i \models \phi$ iff $w_i \models \phi$.
\end{lemma}
\begin{proof}
  Let $Z \subseteq W \times W$ be the relation containing the identity and all $(w_i, v_i)$ pairs:
  \[ Z := \{ (x,x) \mid x \in W \} \cup \{ (w_i, v_i) \mid i \in \Nats \} \]
  We now claim that $Z$ is a bisimulation for the vocabulary $\{a,p\}$.
  This suffices to show that $v_i$ and $w_i$ agree on all $\{a,p\}$-formulas.
  We omit the general definition of bisimulation here and refer to~\cite[Section~2.2]{BRV}.
  \begin{itemize}
  \item Base condition: immediate for the identity pairs, and also holds for the $(w_i, v_i)$ pairs as they agree on $p$
    (and we do not care about $q$ here).
  \item Forth condition:
    For the identity pairs $(x,x)$ this is trivial, so take any $(w_i, v_i)$.
    If $i=0$ then both states have no successors so the condition holds vacuously.
    For $i \neq 0$ the only successors of $w_i$ are $w_{i-1}$ and $u$.
    But we also have $v_i \rel{a} v_{i-1}$ and $v_i \rel{a} u$ so the condition is fulfilled.
  \item Back condition:
    Symmetric to the Forth condition.
    \qedhere
  \end{itemize}
\end{proof}

We now get to the key lemma that will be used to show that there is no interpolant.
In particular we will later choose $j$ to be the modal depth of a hypothetical interpolant.

\begin{lemma}[\lean{depth_equiv}]\label{l:depthEquiv}
  In $\MG$, for any $j \in \Nats$ we have $m \equiv_j^{\{a,p\}} v_j$.
  That is, $m$ and $v_j$ satisfy the same test-free $\{a,p\}$-formulas up to modal depth $j$.
\end{lemma}
\begin{proof}
  To make our induction go through we will prove the following stronger claim:
  In $\MG$ for any test-free $\{a,p\}$-formula $\phi$,
  for any $k$ such that $\depth(\phi) \leq k$ we have $m \models \phi$ iff $v_k \models \phi$.
  That is, for any $k$ ranging from $\depth(\phi)$ onwards the states $m$ and $v_k$ agree on $\phi$.
  Note that choosing $k = j$ implies the lemma statement, so proving this claim suffices.

  To prove the claim, fix any test-free formula $\phi$ such that $\voc(\phi) \subseteq \{a,p\}$.
  We proceed by an induction on the structure of $\phi$.
  As $\phi$ is test-free we do not need mutual induction: the formula may contain programs, but those programs contain no formulas.
  In all cases we take an arbitrary $k$ such that $\depth(\phi) \leq k$.
  \begin{itemize}
  \item The $\bot$ case is trivial.
  \item The only atomic case is $\phi = p$ and $m$ and $v_k$ both make $p$ true.
  \item Cases for $\lnot$ and $\land$ follow from the induction hypothesis.
  \item Case $[\alpha]\psi$.
    Note that $\psi$ has modal depth at most $k-1$.
    We need to show that $m \models [\alpha]\psi$ iff $v_k \models [\alpha]\psi$.\\
    By \autoref{l:boxIffLengths} it suffices to show:
    \begin{center}
      for all $n \in L(\alpha)$ and for all $x$ with $m \rel{a^n} x$ we have $x \models \psi$,\\[0.1em]
      iff\\[0.1em]
      for all $n \in L(\alpha)$ and for all $x'$ with $v_k \rel{a^n} x'$ we have $x' \models \psi$.
    \end{center}
    By first-order reasoning for this it suffices to show:
    \begin{center}
      for all $n \in L(\alpha)$ and all $x$ with $m \rel{a^n} x$ there is an $x'$
      such that $v_k \rel{a^n} x'$ and $x \models \psi \iff x' \models \psi$,\\[0.1em]
      and\\[0.1em]
      for all $n \in L(\alpha)$ and all $x'$ with $v_k \rel{a^n} x'$ there is an $x$
      such that $m \rel{a^n} x$ and $x \models \psi \iff x' \models \psi$.
    \end{center}
    To show these, we provide an overview which sets of states are reachable in $n$ many $a$-steps.
    (To verify this table, consult \autoref{fig:model}).
    \[
      \begin{array}{lll}
        n          & \text{from } m                   & \text{from } v_k                          \\
        \hline
        0          & \{ m \}                          & \{ v_k \}                                 \\
        1          & \{ m, u \}                       & \{ u, v_{k-1} \}                          \\
        2 \ldots k & \{ m, u \} \cup \text{all $w_i$} & \{ m, u, v_{k-n} \} \cup \text{all $w_i$} \\
        k+1 \ldots & \{ m, u \} \cup \text{all $w_i$} & \{ m, u \} \cup \text{all $w_i$}          \\
      \end{array}
    \]
    The remaining claims say that in each row of the table we can ``pair up''
    each state $x$ in the $m$-column with a state $x'$ in the $v_k$-column,
    such that $x$ and $x'$ agree on $\psi$, and vice versa.

    \begin{itemize}
    \item To go from the $m$-column to the $v_k$-column we distinguish cases of $n$ as follows.

      For $n=0$ note that the induction hypothesis with $k$ gives us $m \models \psi$ iff $v_k \models \psi$, exactly what we need.
      (Here we use that $k$ is allowed to be larger than the modal depth of $\psi$.)

      For $n=1$ note that the modal depth of $\psi$ is at most $k-1$,
      so we can apply the induction hypothesis with $k-1$ and get $m \models \psi$ iff $v_{k-1} \models \psi$.
      The only other state reachable in $n=1$ step is $u$ which we can pair with itself.

      For $n \in \{2,\ldots,k\}$ any state that is reachable from $m$ in $n$ steps is also reachable from $v_k$ in $n$ steps so we can pair it with itself.

      Lastly, for $n > k$ we have the same set of reachable states anyway.
    \item
      To go from the $v_k$-column to the $m$-column,
      first note that any non-$v$ state reached from $v_k$ in $n$ steps is also reached from $m$ in $n$ steps, so we can pair each non-$v$ state with itself.
      It remains to pair the $v$ states.

      For $n=0$, we again pair $v_k$ with $m$ and use the induction hypothesis for $k$ as before.

      For $n=1$, we pair $v_{k-1}$ with $m$.
      By the induction hypothesis with $k-1$ we have that this pair also agrees on $\psi$.

      For $n \in \{2,\ldots,k\}$, consider $v_{k-n}$ which is reached from $v_k$ in $n$ steps.
      From $m$ we can reach $w_{k-n}$ in $n$ steps by using the loop at $m$ exactly $n-2$ (possibly 0) times followed by $2$ steps via $u$:
      \[ m \rel{a} \ldots \rel{a} m \rel{a} u \rel{a} w_{k-n} \]
      This chain has length $n - 2 + 2 = n$.
      We can thus pair $v_{k-n}$ with $w_{k-n}$,
      and by \autoref{l:chainBisim} these states agree on $\psi$.

      For $n > k$ there is no $v$-state to be paired, so we are done.
      \qedhere
    \end{itemize}
  \end{itemize}
\end{proof}

\begin{lemma}[\lean{no_interpolant}]\label{lem:noIPexists}
  There is no test-free interpolant for $\bigwedge\Gamma_1 \to \lnot\bigwedge\Gamma_2$.
\end{lemma}
\begin{proof}
  We proceed as outlined in \autoref{sec:overview}.
  Suppose there is an interpolant $\theta$ for $\bigwedge\Gamma_1 \to \lnot\bigwedge\Gamma_2$, and moreover that $\theta$ is test-free.
  Then $\Gamma_1 \cup \{\lnot \theta\}$ and $\Gamma_2 \cup \{\theta\}$ are both not satisfiable.
  Let $j := \depth(\theta)$ and consider the model $\MG$.
  By \autoref{l:modelSatGammaOneTwo} we have $\MG, m \models \Gamma_1$ and $\MG, v_j \models \Gamma_2$.
  By \autoref{l:depthEquiv} in $\MG$ the states $m$ and $v_j$ must agree on $\theta$, because we have $\voc(\theta) \subseteq \voc(\Gamma_1) \cap \voc(\Gamma_2) = \{a,p\}$.
  But then we must have either $m \models \Gamma_1 \cup \{\lnot \theta\}$ or $v_j \models \Gamma_2 \cup \{\theta\}$, contradiction!
\end{proof}

From \autoref{lem:noIPexists} we immediately get the following.

\begin{theorem}[\lean{testFree_no_interpolation}]
  Test-free PDL does not have Craig interpolation.
\end{theorem}

We end this section with an additional note.
In $\Gamma_1$ and $\Gamma_2$ we use only one atomic program and the Kleene star, hence our example shows also that any fragment of PDL that contains at least one atomic program and its Kleene star but not tests cannot have interpolation.
Said differently, it does not matter for our result whether we have the operators `$\cup$' and `${;}$' in the language.
Hence also basic modal logic with Kleene star where we only have $[a]$ and $[a^*]$ as modalities, does not have interpolation.
The common knowledge operator (usually without reflexivity) can be defined using Kleene star; our result thus overlaps with the result of~\cite{StuderCKnoBeth2009}.
A key difference, also discussed by~\cite{StuderCKnoBeth2009} already, is that the common knowledge operator cannot count steps whereas PDL can count parity with programs such as ${(a;a)}^*$.

\section{Methodology}\label{sec:methodology}

The result was obtained in September 2026 using Lean and two LLM systems.
The prompts and summaries of the model replies are available at \url{https://doi.org/10.5281/zenodo.23987961}.
For the Lean formalization, see \url{https://malv.in/2026/test-free-pdl-no-interpolation/}.

We started by naively removing the test operator from the PDL syntax definition in the Lean formalization~\cite{LeanPDLhasCI2026}.
We then ``repaired'' most of the Lean code, e.g.\ by removing cases in proofs and deleting unused code concerning tests.
This repair we did partly ``by hand'' and partly using Aristotle~\cite{achim2025aristotle}.

As expected, we eventually ran into a piece of Lean code that could not easily be repaired:
Definition~9.18 in~\cite{PDLhasCI2026} uses the test operator to define a pre-interpolant.
The construction there uses $[\tau?]\phi$ to express the disjunction $\lnot \tau \lor \phi$.
The motivation to avoid disjunction is that, later, to define actual interpolants from pre-interpolants a fixpoint equation has to be solved.
For that to be possible inside PDL (seen as a fragment of $\mu$-calculus) the pre-interpolants must stay in a certain normal form.
We refer to~\cite[p.~80]{PDLhasCI2026} for details.

We then prompted both Aristotle and Claude Fable to explore alternative constructions, e.g.\ by using a different grammar for pre-interpolants.
This did not lead anywhere, so we eventually also prompted Claude Fable to use the failed proof ideas as inspiration for the opposite goal, namely to find a counterexample.
It did so within a few minutes, including a Lean formalization of the result.
The example we present here is a simplification of the first example found by Claude, using fewer proposition letters and a less complicated Kripke model.
Still, the main idea is the same.

All text in this paper was written by its human author.
The proof structure and lemma statements are strongly inspired by the Lean proof and its comments that were written by Claude.
Also \autoref{fig:unsat} and \autoref{fig:model} were made by prompting Claude to write tikz code.

\paragraph*{Acknowledgements.}
This work was supported by the European Union’s Horizon Europe research and
innovation programme under the Marie Skłodowska-Curie Staff Exchanges action,
grant agreement No 101299559 (SemPER).

I thank the co-authors of~\cite{PDLhasCI2026} and the contributors to~\cite{LeanPDLhasCI2026}, both forming the foundation for finding the example presented here.
I thank
Johan van Benthem,
Balder ten Cate,
Chenwei Shi,
Yde Venema,
Rujing Wang,
Yanjing Wang
and
Junhua Yu
for their helpful suggestions.

\printbibliography[title=Bibliography]

@article{FL:PDL1979,
  author  = {Fischer, Michael J. and Ladner, Richard E.},
  title   = {Propositional dynamic logic of regular programs},
  journal = {Journal of Computer and System Sciences},
  volume  = {18},
  number  = {2},
  pages   = {194--211},
  year    = 1979,
  doi     = {10.1016/0022-0000(79)90046-1}
}

@article{BerPat1981:testFreeWeaker,
  author  = {Francine Berman and Michael Paterson},
  title   = {Propositional dynamic logic is weaker without tests},
  journal = {Theoretical Computer Science},
  volume  = 16,
  number  = 3,
  pages   = {321--328},
  year    = 1981,
  issn    = {0304-3975},
  doi     = {10.1016/0304-3975(81)90102-X}
}

@mastersthesis{MB1988,
  author = {Manfred Borzechowski},
  title  = {{Tableau-Kalkül für PDL und Interpolation}},
  year   = 1988,
  institution = {FU Berlin},
  note   = {Translation by Malvin Gattinger (2020).},
  url    = {https://malv.in/2020/borzechowski-pdl/},
  type   = {Diplomarbeit},
}

@book{Kracht1999tools,
  title     = {Tools and Techniques in Modal Logic},
  url       = {https://wwwhomes.uni-bielefeld.de/mkracht/html/tools/book.pdf},
  author    = {Kracht, Marcus},
  year      = {1999},
  publisher = {Elsevier},
  series    = {Studies in Logic and the Foundations of Mathematics},
  volume    = {142},

}

@book{BRV,
  title     = {Modal Logic},
  series    = {Cambridge Tracts in Theoretical Computer Science},
  number    = {53},
  publisher = {Cambridge University Press},
  author    = {Blackburn, Patrick and Rijke, Maarten de and Venema, Yde},
  year      = {2001}
}

@article{PDLhasCI2026,
  title={Propositional Dynamic Logic has Craig Interpolation: a tableau-based proof},
  author={Manfred Borzechowski and Malvin Gattinger and Helle Hvid Hansen and Revantha Ramanayake and Francisco Trucco Dalmas and Yde Venema},
  journal={Logical Methods in Computer Science},
  year=2026,
  note={Accepted},
  url={https://arxiv.org/abs/2503.13276},
}

@misc{LeanPDLhasCI2026,
  title = {Propositional Dynamic Logic has Craig Interpolation, in Lean 4},
  author = {Malvin Gattinger
            and Amos Nicodemus
            and Djanira dos Santos Gomes
            and Noam Cohen
            and Madeleine Gignoux
            and Wietse Bosman
            and Dionysis Yusofy
            and Haitian Wang
            and Eshel Yaron
            and Xiaoshuang Yang
            and Jeremy Sorkin},
  url = {https://github.com/m4lvin/lean4-pdl},
  doi = {10.5281/zenodo.23053960},
  date = 2026,
}

@misc{achim2025aristotle,
  title         = {Aristotle: {IMO}-level Automated Theorem Proving},
  author        = {Tudor Achim and Alex Best and Alberto Bietti and Kevin Der
                   and Mathïs Fédérico and Sergei Gukov
                   and Daniel Halpern-Leistner and Kirsten Henningsgard
                   and Yury Kudryashov and Alexander Meiburg
                   and Martin Michelsen and Riley Patterson
                   and Eric Rodriguez and Laura Scharff
                   and Vikram Shanker and Vladmir Sicca
                   and Hari Sowrirajan and Aidan Swope and Matyas Tamas
                   and Vlad Tenev and Jonathan Thomm
                   and Harold Williams and Lawrence Wu},
  year          = {2025},
  eprint        = {2510.01346},
  archivePrefix = {arXiv},
  primaryClass  = {cs.AI},
  url           = {https://arxiv.org/abs/2510.01346}
}

@article{StuderCKnoBeth2009,
	title = {Common knowledge does not have the Beth property},
	volume = {109},
	issn = {0020-0190},
	doi = {10.1016/j.ipl.2009.02.011},
	pages = {611--614},
	number = {12},
	journal = {Information Processing Letters},
	author = {Studer, Thomas},
	year = 2009,
}

@article{MaksimovaNextNoBeth1991,
	title = {Temporal logics of ``the next'' do not have the beth property},
	volume = 1,
	issn = {1166-3081},
	doi = {10.1080/11663081.1991.10510772},
	pages = {73--76},
	number = 1,
	journal = {Journal of Applied Non-Classical Logics},
	publisher = {Taylor \& Francis},
	author = {Maksimova, Larisa},
	year = 1991,
}

@inproceedings{GheerCateLTLcraig2009,
	title = {Craig Interpolation for Linear Temporal Languages},
	isbn = {978-3-642-04027-6},
	doi = {10.1007/978-3-642-04027-6_22},
	pages = {287--301},
	booktitle = {Computer Science Logic},
	publisher = {Springer},
	author = {Gheerbrant, Amélie and ten Cate, Balder},
	editor = {Grädel, Erich and Kahle, Reinhard},
	year = {2009},
        series = {Lecture Notes in Computer Science},
        volume = {5771},
}

@book{Kozen1997,
  author    = {Dexter C. Kozen},
  title     = {Automata and Computability},
  publisher = {Springer},
  series    = {Undergraduate Texts in Computer Science},
  year      = {1997},
  doi       = {10.1007/978-1-4612-1844-9}
}

@article{Parikh1966,
  author  = {Rohit J. Parikh},
  title   = {On Context-Free Languages},
  journal = {Journal of the ACM},
  volume  = {13},
  number  = {4},
  pages   = {570--581},
  year    = {1966},
  doi     = {10.1145/321356.321364}
}

@article{GinsburgSpanier1966,
  author  = {Seymour Ginsburg and Edwin H. Spanier},
  title   = {Semigroups, {P}resburger formulas, and languages},
  journal = {Pacific Journal of Mathematics},
  volume  = {16},
  number  = {2},
  pages   = {285--296},
  year    = {1966},
  doi     = {10.2140/pjm.1966.16.285}
}

\end{document}